\documentstyle[11pt,newpasp,twoside,epsf]{article}
\def\edcomment#1{\iffalse\marginpar{\raggedright\sl#1\/}\else\relax\fi}
\reversemarginpar

\begin{document}

\title{
Constraints on the $\Omega_M - \Omega_{\Lambda}$-plane 
from Elliptical Galaxy Counts ?}
\author{S.P.Driver}
\affil{Research School of Astronomy \& Astrophysics, Australia National
University, Canberra, ACT 2600, AUSTRALIA}

\begin{abstract}
Here we present a pilot study into whether elliptical galaxy counts alone,
can place a useful constraint on the $\Omega_M - \Omega_{\Lambda}$-plane. 
The elliptical galaxy counts are drawn from three surveys: The 
Millennium Galaxy Catalogue ($16 > B_{KPNO} > 20$), the B-band Parallel 
Survey ($20 > B_{AB} >24$) and the Hubble Deep Fields ($23 > B_{AB} > 28$). 
The elliptical luminosity function used in the modeling was 
derived from a combination of the Millennium Galaxy Catalogue, the 
two-degree field galaxy redshift survey and the Sloan Digital Sky Survey
($M_{*}^{E/S0}=-19.90, \phi_{*}^{E/S0}=0.0019$ Mpc$^{-3}$ and 
$\alpha^{E/S0}=-0.75$ for $H_{o}=75$km/s/Mpc). 
We adopt a benchmark model and tweak the various input parameters by their
uncertainties to determine the impact upon the counts. We find that {\it if}
the faint-end slope of the 
elliptical galaxy luminosity function is known to $\Delta \alpha < 0.1$,
then over the magnitude range $16 < B < 23$ the counts depend most critically 
upon the cosmology, and can be used to place a weak constraint
on the $\Omega_{M}-\Omega_{\Lambda}$-plane. 
\end{abstract}

\section{Introduction}
Galaxy number-counts as a cosmological probe have been fraught with
difficulties since the conception of the idea by Edwin Hubble in the
1930s (Hubble 1936).  In a recent review Sandage (1997) provides an
insightful historical overview of this topic.  The concept itself was
severely challenged by the work of Tinsley (1977) who showed that the
faint galaxy number-counts depend more critically upon evolution than
upon the cosmological model (at this time only zero-$\Lambda$
models were being considered). In the 1990s even this was superseded by
the faint blue galaxy problem (see Ellis 1997 for a
recent review). It is therefore fair to say that in the lead up to the turn 
of the Millennium 
the use of galaxy number-counts as a cosmological probe was discredited. 
Nevertheless attempts were made and retrospectively may have provided 
the first tentative evidence for a positive cosmological constant 
(Yoshi \& Peterson 1995).

Three factors make the possibility of a revival of galaxy number-counts 
credible, these are: morphological segregation of the faint galaxy population 
and in particular the extraction of ellipticals; advances in our understanding 
of the evolution (or rather non-evolution) of ellipticals (see summary by 
Peebles in these proceedings); and the advent
of $\Lambda$ which dramatically broadens the impact of the cosmology 
upon the counts.

\section{The Data}
Fig. 1 shows the published elliptical count data drawn from three
surveys. The Millennium Galaxy Survey (MGC; Lemon et al 2002); 
the B-band parallel survey (BBPAR; Cohen et al 2002) and the Hubble Deep
Fields (HDFs; Driver et al 1998). The MGC is a ground-based survey over
35 sq degrees along the equatorial strip, conducted at the Isaac Newton 
Telescope using the Wide Field Camera (see Lemon et al 2002).
The B-band parallel survey consists
of WFPC2 F450W observations drawn from the HST archive spanning $\sim 0.04$ sq
degrees (Cohen et al 2002). Finally the deepest data comprises the
Hubble Deep Fields and the deep field 53W02 (Driver
et al 1998) spanning $\sim 0.007$ sq degrees. While these surveys are 
independent bodies of work the techniques applied are not. In particular 
the morphological classifications have been made using the identical ANN 
classifier along with the same set of eyeballs to verify the accuracy. 
This is crucial as it is vitally important for the classification
criterion to be uniform across the entire magnitude range.

\begin{figure}
\plotfiddle{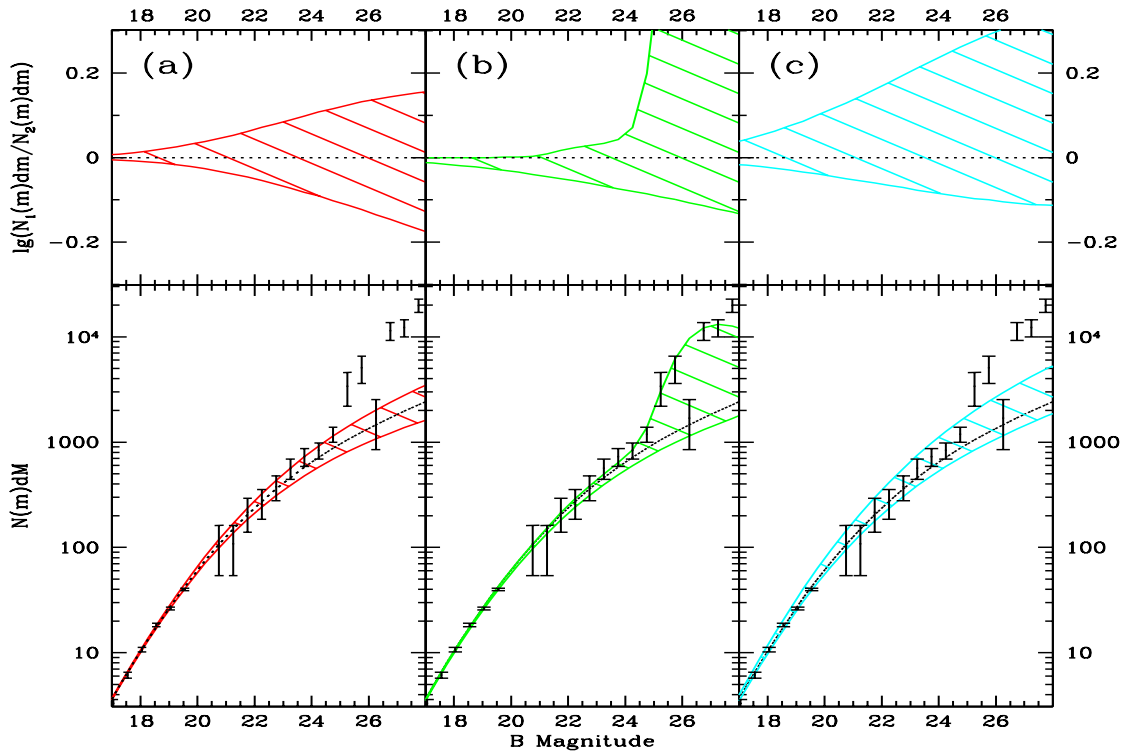}{9.0cm}{0}{75}{55}{-230}{-100}
\caption{
The three panels show elliptical galaxy number-count data and models (lower) 
and the same number-count models with the 
[$\Omega_M=0.3,\Omega_{\Lambda}=0.0$]-benchmark model divided out
(upper). (a) shows the sensitivity of the models to variations in the 
cosmology ([$\Omega_{M}=0.3, \Omega_{\Lambda}=0.7]$, to  
[$\Omega_{M}=1.0, \Omega_{\Lambda}=0.0$]), 
(b) the sensitivity to the evolution 
([$e(z)=(1+z), n(z)=0$] to [$e(z)=0, n(z)=(1+z)^{-0.3}$]) 
and (c) the sensitivity to the faint-end slope
[$\alpha=-0.5$] to [$\alpha=-1.0$]).}
\end{figure}

\section{Constraining $\Omega_{\Lambda}-\Omega_{Matter}$}
To model the elliptical counts we require 8 parameters ($\phi_*, M_*,
\alpha, e(z), n(z), k(z), \Omega_{M}, \Omega_{\Lambda}$), by marginalising
over $\phi_*$ and solving for $\Omega_{M}-\Omega_{\Lambda}$ this requires
5 known parameters. Here we adopt as our benchmark model the following
($H_{o}=75$km/s/Mpc):

~

\noindent
$M_*=-19.9, \alpha=-0.75, e(z)=0, n(z)=0$,
$k(z)=4z+1.5z^2-1.3z^3+0.27z^4$

~

\begin{figure}
\plotfiddle{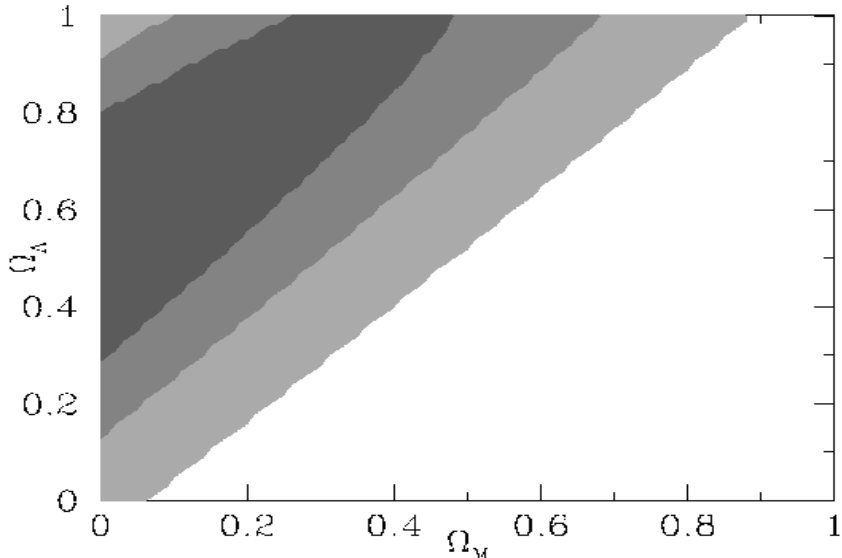}{6.5cm}{0}{60}{50}{-200}{-100}
\caption{The 1-,2- and 3-$\sigma ~ \chi^2$-contours in the
$\Omega_{M}-\Omega_{\Lambda}$-plane for the standard benchmark 
model (see text), i.e., zero net evolution}
\end{figure}

\noindent
Fig. 1 shows the assembled galaxy number-counts along with our benchmark 
model (dotted line). We now perturb the benchmark model as indicated 
in the figure caption to show how the models depend upon: Fig.1(a), the cosmology; Fig.1(b),
the evolutionary model (pure-luminosity or number-density); and Fig.1(c),
the faint-end slope of the local luminosity function. The upper panels show
the same with the benchmark model divided out. At very faint magnitudes 
($B > 24$) mags the uncertainty in the evolution entirely dominates the
galaxy counts. At intermediate and bright magnitudes it is the uncertainty in
$\alpha$ which dominates. However pressing further we find that 
if $\Delta \alpha < 0.1$ then it is the cosmology which dominate the counts 
over the range $16 < B < 23$ mags. It should therefore be viable to place a 
constraint on the $\Omega_{M}-\Omega_{\Lambda}$-plane using {\it bright} 
morphological galaxy counts. Fig. 2 shows the resulting $\chi^2$ 1-, 2- and 3-
$\sigma$ regions in the 
$\Omega_M-\Omega_{\Lambda}$-plane for the benchmark model (i.e.,
$M_*$, $\alpha$, $k(z),e(z)$ and $n(z)$ fixed to the values above) and 
marginalising over $\phi_*$. 
The benchmark counts clearly favour a non-zero cosmological 
constant but how much is this due to the adoption of the five fixed 
parameters. 
Marginalising over all parameters requires extensive supercomputer time
currently not available, however in Fig. 3 we show the results for three
perturbations of our benchmark model to indicate the likely impact of some of
these parameters upon the results. These perturbations to the benchmark model 
are shown below:

~

\noindent
Perturbation 1 (Fig 3a): $e(z)=(1+z)^1$, a pure-luminosity evolution model

\noindent
Perturbation 2 (Fig 3b): $n(z)=(1+z)^{-0.3}$, a number-density evolution model

\noindent
Perturbation 3 (Fig 3c): $\alpha=-1.0$, an uncertainty in $\alpha$

~

\noindent
It can be seen that the change from pure-luminosity evolution, 
to zero-evolution, to number-density evolution mimics $\Lambda$. I.e., if
galaxies were brighter in the past one needs to reduce the volume element in 
order to
keep the galaxy-counts constant etc. Curiously neither extreme allows for an 
$\Omega_M=1$ universe. More worrisome however is the critical dependency on 
$\alpha$. Current published values range from $\alpha=-0.5$ to $\alpha=-1$
and until it is pinned down an $\Omega_{M}=1$ universe is allowed.
Hence our conclusion is that it is the 
uncertainty in the local luminosity function parameters and in particular 
$\alpha$ which currently prevents a serious credible constraint on the 
$\Omega_{M}-\Omega_{\Lambda}$-plane.

\begin{figure}
\plotfiddle{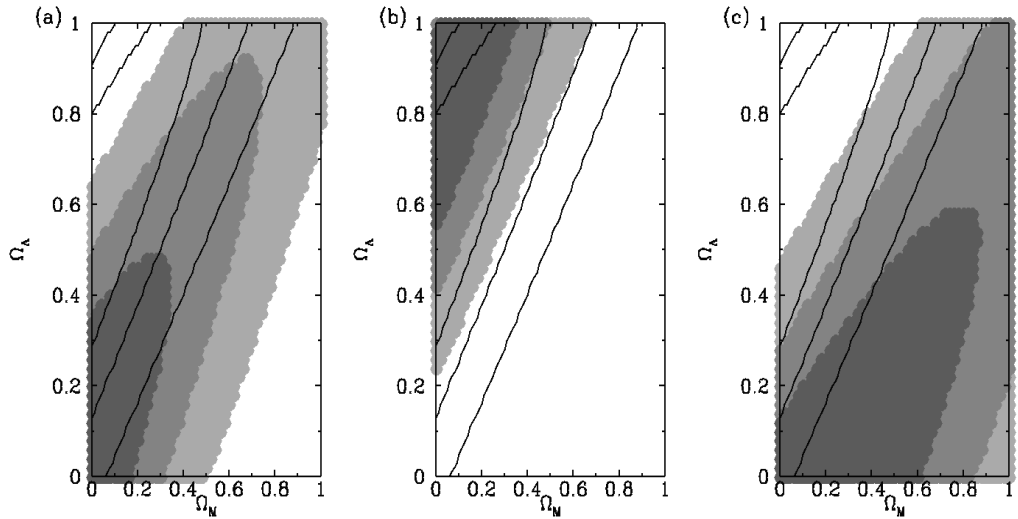}{6.5cm}{0}{75}{85}{-220}{-240}
\caption{The $\Omega_{M}-\Omega_{\Lambda}$-plane for our three perturbations 
of the benchmark model (see text). Lines indicate contours from Fig 2.}
\end{figure}

\end{document}